\documentstyle[multicol,epsf,aps,pre]{revtex}
\begin{document}
\title{Dissolution in a field}
\author{W. Hwang$^\dagger$ and S. Redner}
\address{Center for BioDynamics, Center for Polymer Studies, 
and Department of Physics, Boston University, Boston, MA 02215}

\maketitle 
\begin{abstract} 
  We study the dissolution of a solid by continuous injection of reactive
  ``acid'' particles at a single point, with the reactive particles
  undergoing biased diffusion in the dissolved region.  When acid encounters
  the substrate material, both an acid particle and a unit of the material
  disappear.  We find that the lengths of the dissolved cavity parallel and
  perpendicular to the bias grow as $t^{2/(d+1)}$ and $t^{1/(d+1)}$,
  respectively, in $d$-dimensions, while the number of reactive particles
  within the cavity grows as $t^{2/(d+1)}$.  We also obtain the exact density
  profile of the reactive particles and the relation between this profile and
  the motion of the dissolution boundary.  The extension to variable acid
  strength is also discussed.

\end{abstract}
\begin{multicols}{2}
\narrowtext

\section{Introduction}

The dissolution of a solid material by contact with a reactive fluid is a
fundamental process that underlies corrosion\cite{corr},
diagenesis\cite{daccord,sahimi}, erosion\cite{meng},
etching\cite{kim,sapoval}, and many other industrial processes.  The same
dynamical process can also be viewed as the melting of a solid by heating the
material at a single point in the interior\cite{melt}.  These types of
dissolution (or melting) processes are described by the motion of the
interface between the reactive fluid and the solid.  In many situations,
molecular diffusion is the transport mechanism for the reactive particles,
and this leads to diffusion-controlled moving boundary value
problems\cite{mov,cummings}.

\begin{figure}
\narrowtext
\epsfxsize=2.8in
\hskip 0.15in\epsfbox{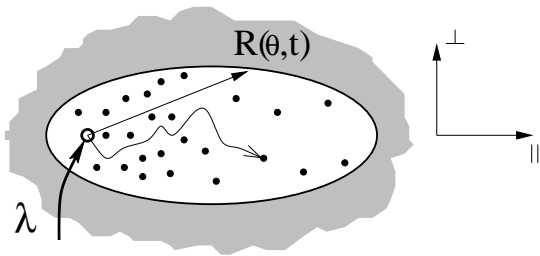}
\vskip 0.1in
\caption{Schematic illustration of the dissolution process.  Reactive
  particles (dots) are continuously injected at rate $\lambda$ at a single
  point (circle).  Each particle undergoes biased diffusion, with bias in the
  parallel direction.  When a particle reaches the boundary of the dissolved
  cavity, a unit of the host material and the particle both disappear.}
\label{cartoon}
\end{figure}

In this work, we consider the kinetics of this dissolution process when there
is a superimposed bias on the diffusive motion of the acid.  Such a bias can
be easily realized, for example, by an electric field acting on ionized
particles, or by a gravitational field or a pressure gradient acting on a
flowing fluid.  We find that the bias is a relevant perturbation with respect
to molecular diffusion and gives rise to a dissolution process different from
that caused by isotropic diffusion\cite{larralde}.  As a function of time,
the size of the dissolved region grows continuously and preferentially in the
direction of the bias (Fig.~\ref{cartoon}).  The basic questions that we
shall study are the density profile of the acid particles inside the
dissolved cavity, as well as the shape and time dependence of the boundary
between the fluid and unreacted material.

In Sec.\ II, we first define the model and write the reaction diffusion
equation that governs the density of reactive particles in the continuum
limit.  In Sec.\ III, we then solve for the steady state density profile of
reactive particles in the dissolved cavity.  This profile satisfies the
anisotropic Laplace equation, which is the time-independent limit of the
basic equation of motion.  In Sec.\ IV, we investigate the motion of the
interface and determine the different characteristic lengths of the cavity in
the directions parallel and perpendicular to the bias.  We briefly summarize
in Sec.\ V and also discuss a generalization of the system to variable acid
strength.

\section{The model}

We consider the following microscopic dissolution process
(Fig.~\ref{cartoon}).  Initially all sites are unreacted.  Acid particles are
injected at rate $\lambda$ at the origin of a $d$-dimensional solid
substrate.  After injection, an acid particle undergoes biased diffusion
until it hits a site at the interface between unreacted substrate and the
dissolved cavity.  In this interaction, both the host substrate site and the
acid disappear.  We can think of the acid as having unit strength so that one
acid particle and one substrate particle are consumed in a reaction.  Later
we will generalize to allow acid to dissolve many substrate sites before
being neutralized.  In the context of melting, we can think of particle
injection as the localized input of heat and dissolution as the melting of
the solid when heat reaches the interface.

In the limiting case where the reactive particles undergo isotropic
diffusion, the resulting dissolution process has been extensively studied,
both in the context of melting\cite{melt} and in the framework of
diffusion-controlled reactions\cite{larralde}.  Here the radius of the
dissolved region $R(t)$ grows as $(t\ln t)^{1/2}$ and as $t^{1/d}$ for
spatial dimension $d=1$ and $d\geq2$, respectively.  The density profile of
the acid particles is also radially symmetric and asymptotically approaches a
steady state as a function of the scaled radial distance $r/R(t)$.

These results have a simple origin.  In $d=1$, for $N$ diffusing particles
initially located at the origin, the furthest particle from the origin after
time $t$ will be a distance of the order of $(t\ln N)^{1/2}$\cite{extreme}.
Thus if $\lambda t$ acid particles are injected continuously at the origin,
the most distant particle, and therefore the position of the interface should
be $(t\ln\lambda t)^{1/2}$ from the origin.  For $d\ge 2$, since each acid
particle dissolves a single substrate particle, the dissolved volume can be
at most $\lambda t$.  Consequently, the radius can be no larger than
$t^{1/d}$ for $d\ge2$.  Within the dissolved cavity, the density profile of
acid particles away from the interface approaches a static limit for $d>2$.
This density profile thus obeys the Laplace equation and decays as $r^{2-d}$.
For $d\leq 2$ the density profile is not static and it can be obtained
conveniently by a scaling solution\cite{larralde}.

In the presence of the bias, we need to consider separately the growth
parallel and perpendicular to the bias.  The dynamics of this anisotropic
dissolution process is governed by $c(\vec r,t)$, the concentration of acid
at position $\vec r$ within the dissolved region at time $t$.  This
concentration obeys the convection-diffusion equation
\begin{equation}
\label{cdeq}
{\partial c\over \partial t}+v{\partial c\over\partial x} = 
D\nabla^2c +\lambda\delta(\vec r),
\end{equation}
subject to the absorbing boundary condition $c(\vec r,t)=0$ for $|\vec
r|=R(\theta,t)$, where $R(\theta,t)$ is the radius of the moving interface as
a function of $\theta$ and $t$ (Fig.~\ref{cartoon}).  Here we have taken the
bias direction as along $x$.  The motion of the interface is then governed by
the flux of acid onto the interface
\begin{equation}
\label{bc}
{\partial\vec R\over\partial t}=-KD\vec\nabla c|_{|\vec r|=R(\theta,t)},
\end{equation}
where $K$ is the parameter which quantifies the acid strength.  Here we
define this constant to be 1 and later generalize to arbitrary acidity.
Notice also that there is no convective contribution to this flux ($vc$)
because of the absorbing boundary condition.

A basic feature which simplifies much of the analysis is that the density
profile of the acid within the dissolved region is stationary in time except
near the boundary.  This arises because the input of new particles
compensates for their loss at the boundary.  This same simplifying feature,
which also applies in the case of isotropic diffusion for $d>2$,
ultimately stems from the transient nature of biased diffusion \cite{redner}.
We now exploit this stationarity to obtain the exact concentration profile of
acid within the dissolved cavity.

\section{Steady-state concentration profile}

Setting the time derivative in Eq.~(\ref{cdeq}) equal to zero, an anisotropic
Laplace equation results.  For zero bias, this gives the classical Laplace
equation with steady-state solution $c_{\rm ss}(\vec r)\propto r^{2-d}$ for
$d>2$.  To find the corresponding solution in the presence of a bias, we
perform a Fourier transform of the anisotropic Laplace equation to yield
\begin{equation}
\label{c_ft}
-D\vec k^2\tilde c(\vec k)+ivk_x\tilde c(\vec k)+\lambda=0,
\end{equation}
with solution
\begin{equation}
\label{c_ft-soln}
\tilde c(\vec k)={\lambda\over{D\vec k^2-ivk_x}}.
\end{equation}
Inverting this Fourier transform gives the steady-state acid
concentration
\begin{eqnarray}
\label{c_inv}
c_{\rm ss}(\vec r)&=&\int\!\!{d\vec k\over(2\pi)^d}\,\,\tilde c(\vec k)\,
       e^{-i\vec k\cdot\vec r}\nonumber\\ 
  &=&{\lambda\over(2\pi)^dD}\int\!\!d\vec k\,\,
            {e^{-i\vec k\cdot\vec r}\over \vec k^2-ivk_x/D}\nonumber\\
  &=&{\lambda\over(2\pi)^dD}\int\!\!d \vec k\,\,
            {e^{-i\vec k\cdot\vec r+{vx\over2D}}
           \over k^2+({\vec v\over2D})^2}.
\end{eqnarray}
In the last step, we complete the square in the denominator and then shift
$k_x$ by $k_x-iv/2D$.  The last integral in Eq.~(\ref{c_inv})
is\cite{gradstein},
\begin{equation}
\label{steadysoln}
c_{\rm ss}(\vec r)={\lambda e^{vx\over2D}\over (2\pi D)^{d\over2}} 
   \left({v\over 2r}\right)^{{d\over2}-1}K_{{d\over2}-1}
   \left({vr\over 2D}\right),
\end{equation}
where $K_{d/2-1}$ is the modified Bessel function. 

This exact solution has very different forms in the regions $x>0$ and $x<0$.
In the interesting case of $x\gg 0$, we substitute the asymptotic expansion
$K_\nu(z)\sim (\pi/2z)^{1/2}e^{-z}$ \cite{as} into Eq.~(\ref{steadysoln}) to
obtain
\begin{equation}
\label{cs-asymp}
c_{\rm ss}(\vec r)\sim {\lambda\over v}\left({v\over{4\pi
      Dr}}\right)^{(d-1)/2} e^{-v(r-x)/2D}.
\end{equation}
In the special cases of $d=1$, 2, and 3, this reduces to
\begin{equation}
\label{cs2}
c_{\rm ss}(\vec r)=\cases{{\displaystyle{\lambda\over v}}, & $d=1$\cr\cr
          {\displaystyle{\lambda\over\sqrt{4\pi Dvr}}\,\,e^{-v(r-x)/2D}}, & $d=2$\cr\cr
         {\displaystyle{\lambda\over4\pi Dr}\,\,e^{-v(r-x)/2D}}, & $d=3$.}
\end{equation}
Conversely, for $x<0$, $c_{\rm ss}(\vec r)$ decays exponentially as a
function of the distance from the origin, with the length scale of this decay
proportional to $D/v$.

\section{Interface motion}

To gain a fuller appreciation for time-dependent features and the motion of
the interface, we have performed Monte Carlo simulations of the dissolution
process.  Our simulations are based on simply tracking the motion of all the
reactive particles.  Each particle performs a biased nearest-neighbor random
walk on a $d$-dimensional hypercubic lattice, with hopping probability equal
to $1/(2d)$ in the $2d-2$ directions perpendicular to the bias, and equal to
$1/(2d)<p_+<1/d$ and $p_-=1/d-p_+<p_+$ in the $\pm x$ directions,
respectively.  These hopping probabilities give a bias velocity $v=p_+-p_-$,
as well as a superimposed isotropic diffusion process, with the diffusion
coefficient in all coordinate directions equal to $1/2d$.  We henceforth fix
the injection rate to be $\lambda=1$.  Each lattice site is initially
regarded as one unit of solid material which disappears when it is contacted
by a reactive particle.
  
The choice of the bias in our simulations is dictated by basic physical
considerations.  If the bias velocity is too small, there is a long crossover
time before the bias dominates over the diffusion.  On the other hand, for a
bias velocity which is close to the maximum value of $1/d$, the length of the
dissolved region becomes extremely large and this requires considerable
computer memory to store the data of the system map.  For these reasons, we
found it optimal to consider intermediate values of the velocity.

\begin{figure}
\narrowtext
\epsfxsize=2.9in
\hskip 0.15in\epsfbox{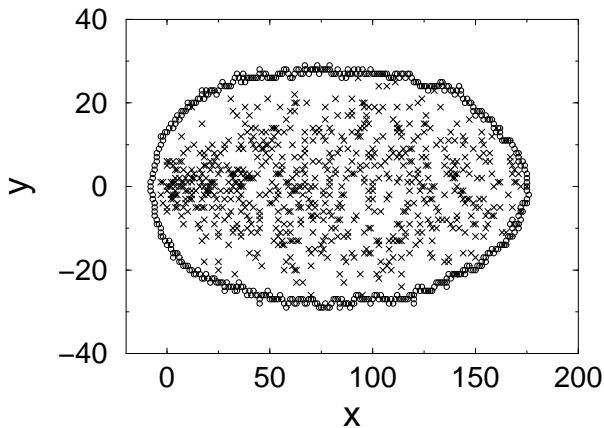}
\vskip 0.2in
\caption{Typical shape of the dissolved region on the square lattice after 
  $t=10^4$ time steps.  Reactive particles and sites on the dissolution
  boundary are denoted by crosses and circles, respectively.  The injection
  point is at $(x,y)=(0,0)$ and the bias velocity is $v=0.2$.  For this
  velocity, $vt\gg\sqrt{2Dt}$ and thus the system is far beyond the initial
  transient regime.}
\label{fig_oval}
\end{figure}

As a function of time an elliptically-shaped dissolved cavity grow in which
the interface remains relatively smooth (Fig.~\ref{fig_oval}).  Within this
cavity, there is a distribution of mobile reactive particles which have not
yet reached the interface.  These physical characteristics have different
dependences in spatial dimension $d=1$ and in higher dimensions; we therefore
discuss these two cases separately.

\subsection{One Dimension}\label{1d}

We first apply a simple flux balance argument\cite{infiltration} to show that
for $x>0$ the interface boundary $R(t)$ moves with a fixed propagation
velocity -- defined to be $v_I$ -- which is less than the particle velocity
$v$.  Since the input of reactive particles occurs at rate $\lambda$, the
particle flux in the $+x$ direction is simply $\lambda$.  Thus a unit flux
would lead to an interface velocity $v_I/\lambda$.  On the other hand, in a
reference frame that moves at velocity $v$, the reactive particles are at
rest while the substrate particles (with density $\rho$) move with velocity
$-v$.  In this moving frame, a flux of substrate particles $-\rho v$ would
lead to the interface moving at velocity $v_I-v$.  Therefore a unit particle
flux would give an interface velocity $(v_I-v)/(-\rho v)$.  Since the
reaction has the symmetrical stoichiometry $A (\rm acid)+B (\rm substrate)\to
0$, the two velocities under conditions of unit flux must be equal.  This
then leads to the interface velocity
\begin{equation}
\label{vint}
v_I=\lambda v/(\lambda+\rho v).
\end{equation}

For $r\approx R$, the density profile decreases sharply from the constant
value $\lambda/v$ (Eq.~(\ref{cs2}) to 0.  Because the dissolution process is
equivalent to the reaction $A+B\rightarrow 0$ with components approaching
each other at finite velocity, the width of the reaction front is
proportional to $D/v$ and does not grow in time\cite{infiltration,br}. We
have also verified these features by Monte Carlo simulation (data not shown).

\subsection{Dimensions $d\geq 2$}

For $d\ge2$, let us locate the reactive particles by the $d$-dimensional
cylindrical coordinates $\vec r=(x,\vec r_\bot)$, where $\vec r_\bot$ is the
$d-1$-dimensional radial vector perpendicular to the $x$ axis.  Similarly, we
write $\vec R=(R_\|, \vec R_\bot)$ to denote the position of the interface.
Since the dissolved cavity grows predominantly along the direction of the
bias, we focus our attention on this downstream portion of the interface.

\begin{figure}
  \narrowtext \epsfxsize=3.0in \hskip 0.05in\epsfbox{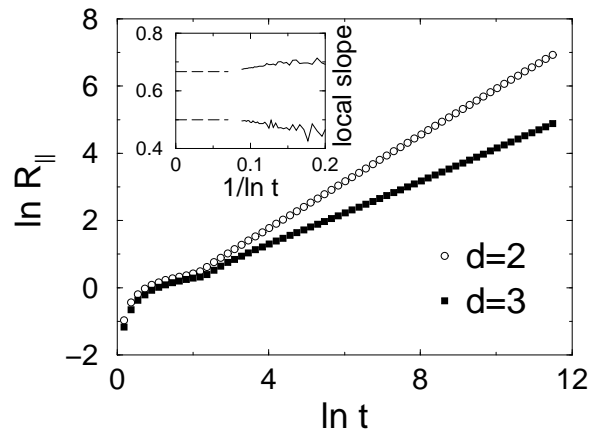}
  \vskip 0.2in
\caption{Plot of $R_\|$ versus $t$ on a double logarithmic scale for
  $d=2$ and 3 (bias velocity $v=0.3$ for both cases).  The data represent
  averages over 500 ($d=2$) and 1000 ($d=3$) realizations.  The inset shows
  the local slopes of the data versus $1/\ln t$.  These appear to converge to
  2/3 in $d=2$ and 1/2 in $d=3$ (dashed lines), in agreement with
  Eq.~(\ref{Rs}).}
\label{fig_r_parall}
\end{figure}

We now determine how $R_\|$ and $R_\perp$ depend on time.  Let us assume that
$R_\|\sim t^{\nu_\|}$ and $R_\perp\sim t^{\nu_\perp}$.  Since the motion of
the reactive particles in the transverse direction is diffusive, we expect
that $R_\perp\propto R_\|^{1/2}\sim t^{\nu_\|/2}$.  Then the volume of the
dissolved region is proportional to $V\sim R_\|R_\perp^{d-1} \sim
t^{(d+1)\nu_\|/2}$.  Since $V$ cannot grow faster than $\lambda t$, we must
have $(d+1)\nu_\|/2 \le 1$.  On the other hand, if $V$ were to grow slower
than $\lambda t$, the number of reactive particles in the dissolved region
would have to grow with time, in contradiction with the steady state density
profile derived above.  Thus $V$ should grow linearly with time, from which
we conclude that $\nu_\|=2/(d+1)$.  Hence
\begin{equation}
\label{Rs}
R_\|\sim t^{2/(d+1)}\qquad R_\perp \sim t^{1/(d+1)}.
\end{equation}
These predictions are in very good agreement with our numerical simulations
(Fig.~\ref{fig_r_parall}).

\begin{figure}
\narrowtext
\epsfxsize=3.0in\hskip 0.05in\epsfbox{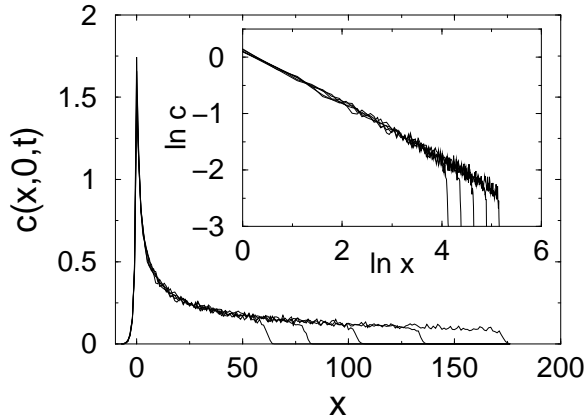}
\vskip 0.2in
\caption{Density profile of reactive particles in $d=2$ along the $x$
  axis at equally spaced times on a logarithmic scale.  The bias velocity
  is $v=0.2$.  The data represent averages over 1500 realizations.  Inset:
  same data on a double logarithmic scale.  Except for the sharply
  decreasing interfacial region, the profile $c_{\rm ss}(x,0)\propto
  x^{-1/2}$ (see Eq.~(\ref{cs2})).}
\label{fig_shape2d}
\end{figure}

The dependence of $R_\|$ on $t$ can also be determined independently from the
density profile $c(\vec r,t)$.  Similar to the case of $d=1$, $c(\vec r,t)$
approaches $c_{\rm ss}$ far from the interface, while $c(\vec r,t)$ assumes a
traveling wave form near the interface, which rapidly decays from $c_{\rm
  ss}$ to 0 (Fig.~\ref{fig_shape2d}).  From the inset to this figure, we see
that the width of this front does not grow in time.  Using Eq.~(\ref{bc}), we
can then approximate the equation of motion for the interface as $\dot{R}\sim
c_{\rm ss}/w$, where $w\sim D/v$ is the width of the front.  Substituting the
asymptotic expansion for $K_\nu(z)$ in Eq.~(\ref{steadysoln}) then gives
$c_{\rm ss}(R_\|)\sim R_\|^{(1-d)/2}$.  Using this in $\dot R\sim c_{\rm
  ss}/w$, we obtain $R_\|\sim \left(t/w\right)^{2/(d+1)}$, in agreement
with Eq.~(\ref{Rs}).

\subsection{Scaling for the density profile}

To obtain the number of reactive particles, it is convenient to express their
density profile in a scaled form.  Based on the time dependences of $R_\|$
and $R_\bot$, we introduce the scaled variables $\xi_\|=x/t^{2/(d+1)}$ and
$\vec\xi_\bot=\vec r_\bot/t^{1/(d+1)}$.  In terms of these scaled
coordinates,
\begin{eqnarray}
\label{r_scale}
r &=& (x^2+{\vec r\/_\perp}^2)^{1/2} \nonumber \\
 &\simeq& t^{2/(d+1)}\xi_\|+{1\over2}{\xi_\bot^2\over\xi_\|}.
\end{eqnarray}
Using the asymptotic expansion of $K_\nu$, and substituting the scaled
variables into Eq.~(\ref{steadysoln}), we obtain the scaling form for the
density profile
\begin{eqnarray}
\label{scalesoln}
c(\xi_\|,\vec\xi_\bot,t) &\sim& t^{-{d-1\over d+1}}\,\,\xi_\|^{-{d\over2}}
       \exp\left({-{v\over 2D}{\xi_\bot^2\over\xi_\|}}\right),\nonumber \\
   &\equiv &  t^{-{d-1\over d+1}}\,\, f(\xi_\|,\xi_\bot).
\end{eqnarray}

\begin{figure}
\narrowtext
\epsfxsize=3.0in\hskip 0.05in\epsfbox{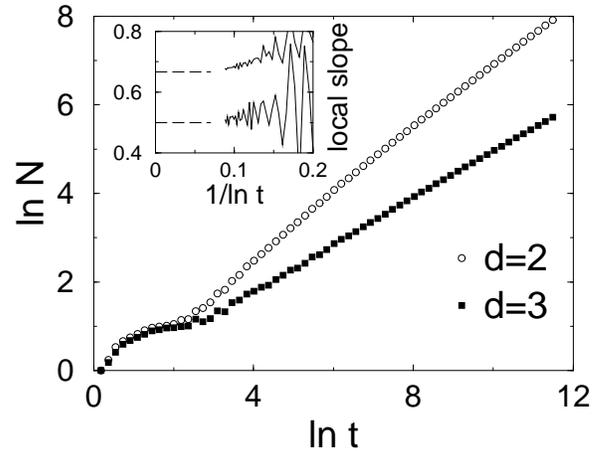}
\vskip 0.2in
\caption{Plot of $\ln N(t)$ versus $\ln t$.  The data are from the same 
  simulations as Fig.~\ref{fig_r_parall}.  The inset shows the local
  slopes which appear to converge to $2/3$ and $1/2$ for $d=2$ and 3 (dashed
  lines), consistent with Eq.~(\ref{Nt}).}
\label{fig_nt}
\end{figure}

{}From this form, we easily obtain the time dependence of the total
number of active particles $N(t)$ to be
\begin{eqnarray}
\label{Nt}
N(t) &=&\int d\vec r \,c(\vec r,t), \nonumber \\
  &\sim &t^{2/(d+1)} \int d\xi_\| d\vec \xi_\perp\, f(\xi_\|,\vec\xi_\bot),
  \nonumber \\
  &\sim & t^{2/(d+1)}.
\end{eqnarray}
This prediction is also in excellent agreement with our simulations
(Fig.~\ref{fig_nt}).

Alternatively, $N(t)$ equals the difference between the number of
injected particles and the volume of the dissolved region.  We use this fact
to provide a more precise form for the time dependence of the dissolved
volume $V(t)$.  By multiplying Eq.~(\ref{bc}) by the surface element $d\vec
S$, integrating over the dissolution interface, and using Eq.~(\ref{cdeq}),
we have
\begin{eqnarray}
\label{surface}
\int d\vec S\cdot {d\vec R\over dt}&=& -D\int d\vec S\cdot
    \vec\nabla c|_{\vec R}\nonumber\\
  &=&-D\int dV\,\,\nabla^2 c\nonumber\\
 &=&-\int dV\,\left(\partial_t c+v\partial_x c -\lambda\delta(\vec
    r)\right)\nonumber\\
 &=& -{dN\over dt}+\lambda.
\end{eqnarray}
The left hand side is simply equal to $\dot{V}$.  Therefore we obtain the
obvious conservation equation ${d\over dt}(V+N)=\lambda$.  Then
Eq.~(\ref{Nt}) gives $V(t)\sim \lambda t-\alpha t^{2/(d+1)}$, where $\alpha$
is a constant related to the integral in Eq.~(\ref{Nt}).

\section{Discussion}

In this paper, we studied the dissolution of a substrate when acid particles
are continuously injected at a single point and there is an external field
which causes these particles to undergo biased diffusion.  The basic
quantities of interest in this process are the concentration profile of the
acid and the growth kinetics of the dissolved region.  Within the dissolved
region, the acid concentration follows the steady state profile of biased
diffusion; this is just the solution of the anisotropic Laplace equation.
The shape of the dissolved region is strongly anisotropic with its length
growing in time as $\xi_\|\sim t^{2/(d+1)}$ while the transverse width grows
as $\xi_\perp\sim t^{1/(d+1)}$.

\begin{figure}
\narrowtext
\hskip 0.35in\epsfxsize=2.8in
\epsfbox{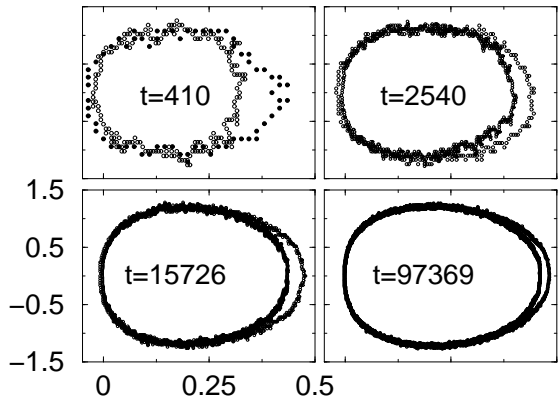}
\vskip 0.4in
\caption{Scaled boundary profiles for acid strength $A=1$ and $A=5$ at different
  times.  All graphs are on the same scale.  The coordinates for $A=5$ are
  divided by $(5t)^\nu$, with $\nu=2/3$ (horizontal scale), $1/3$ (vertical
  scale).  The smaller contours are for $A=5$. }
\label{fig_visited}
\end{figure}

A simple and relevant extension of our model is to the case of variable acid
strength.  This can be realized by assuming a substrate density $\rho>1$ so
that $\rho$ acid particles must hit a given substrate site before it
dissolves (weak acid) or that an acid particle dissolves $A>1$ substrate
sites before becoming neutralized (strong acid).  In the current model,
$\rho=A=1$.  The case $A>1$ is equivalent to having a particle density in the
substrate $\rho$ equal to $1/A$ and with an acid particle dissolving one
substrate particle.  Incorporating this scaling behavior into Eq.~(\ref{bc})
for the interface motion we have
\begin{equation}
\label{bc_general}
{\partial\vec R\over\partial t}=-{D\over\rho}\vec\nabla c|_{|\vec r|=R(\theta,t)}.
\end{equation}

For $\rho>1$, the dissolution boundary becomes smoother and grows more slowly
since it takes many acid particles to dissolve each substrate site.  For
$d=1$, this follows immediately from the expression $v_I=\lambda
v/(\lambda+\rho v)$ which was obtained from the flux balance argument
(Sec.~\ref{1d}).  In general, Eqs.~(\ref{cdeq}) and (\ref{bc}) are invariant
after normalizing $c\to c/\rho$ and rescaling $\lambda\to\lambda/\rho$.  This
means that a change of the substrate density or acid strength will only
change the time scale through the injection rate.  We have tested this
hypothesis by simulations in which acidity $A$ varied between 1 and 160.  At
short times, the dissolution boundary appears to be much rougher for
$\rho<1$.   Asymptotically, it appears that the boundaries for
different values of $\rho$ approach a common limit.  The overall
effect of varying the acidity is simply to change the time scale.  However,
the subdominant terms to Eqs.~(\ref{Rs}) and (\ref{Nt}) seem to have strong
acidity dependence so that there is a long-lived transient correction to this
simple scaling behavior (Fig.~\ref{fig_visited}).

\bigskip {\bf Acknowledgments} We thank Paul Krapivsky for helpful
discussions.  We are also grateful to grants ARO DAAD19-99-1-0173 and NSF
DMR9978902 for financial support.

\end{multicols}

\begin{thebibliography}{99}

\bibitem[\dagger]{cbe} Current address: Center for Biomedical Engineering,
Massachusetts Institute of Technology, Cambridge, MA 02139

\bibitem{corr} H. H. Uhlig, {\em Corrosion and Corrosion Control}, (Wiley,
  New York, 1963). 

\bibitem{daccord} G. Daccord, Phys.\ Rev.\ Lett.\ {\bf58}, 479 (1987)

\bibitem{sahimi} M. Sahimi, G. R. Gavalas, T. T. Tsotsis, Chem.\ Engr.\ Sci.\
  {\bf 45}, 1443 (1990).

\bibitem{meng} H. F. Meng and E. G. D. Cohen, Phys.\ Rev.\ E {\bf 51},
  3417 (1995)

\bibitem{kim} K. S. Kim, J. A. Hurtado, and H. Tan, Phys.\ Rev.\ Lett.\
  {\bf 83}, 3872 (1999)

\bibitem{sapoval} A. Gabrielli, A. Baldassarri, and B. Sapoval, Phys.\ Rev.\
  E {\bf 62}, 3103 (2000).
  
\bibitem{melt} H. S. Carslaw and J. C. Jaeger, {\em Conduction of Heat in
    Solids} (Oxford University Press, New York, 1959).

\bibitem{mov} J. Crank, {\em Free and Moving Boundary Value
    Problems} (Oxford University Press, New York, 1987).
  
\bibitem{cummings} L. M. Cummings, Y. Hohlov, S. D. Howinson, and 
  K. Kornev, J. Fluid Mech.\ {\bf 378}, 1 (1999)

\bibitem{larralde} H. Larralde, Y. Lereah, P. Trunfio, J. Dror,
  S. Havlin, R. Rosenbaum, and H. E. Stanley, Phys.\ Rev.\ Lett.\
  {\bf 70}, 1461 (1993).
  
\bibitem{extreme} This is a simple exercise in extreme value statistics.  For
  a general introduction see {\it e.\ g.}, J. Galambos, {\sl The Asymptotic
    Theory of Extreme Order Statistics} (R. E. Krieger Pub., Malabar, FL,
  1987).  For a discussion of the location of the extreme particle in
  diffusion see S.~Redner and P.~L.~Krapivsky, Am.\ J. Phys.\ {\bf 67},
  1277-1283 (1999)
  
\bibitem{redner} S. Redner, {\em A Guide to First Passage Processes},
(Cambridge University Press, New York, 2001).

\bibitem{gradstein} I. S. Gradstein and I. M. Rhzhik,
 {\em Table of Integrals, Series, and Products}, Academic Press, New
 York (1967).
 
\bibitem{as} {\em Handbook of Mathematical Functions} edited by M. Abramowitz
and I.~A.~Stegun, (Dover, New York, 1965).

\bibitem{infiltration} W. Hwang and S. Redner, Phys.\ Rev.\ E. {\bf63}
  021508 (2001).

\bibitem{br} E. Ben-Naim and S.~Redner, J. Phys.\ A {\bf 25}, L575 (1992). 


\end{thebibliography}
\end{document}